\documentclass[preprint,showpacs,superscriptaddress]{revtex4}

\usepackage{amssymb}
\usepackage{amsmath}
\usepackage{graphicx}
\usepackage{natbib}
\usepackage{nicefrac}
\usepackage{multirow} 
\usepackage{comment}

\begin{document}

\title{Circuit Quantum Electrodynamics with a Spin Qubit}
\author{K. D. Petersson}
\affiliation{Department of Physics, Princeton University, Princeton, NJ 08544, USA}
\author{L. W. McFaul}
\affiliation{Department of Physics, Princeton University, Princeton, NJ 08544, USA}
\author{M. D. Schroer}
\affiliation{Department of Physics, Princeton University, Princeton, NJ 08544, USA}
\author{M. \nolinebreak Jung}
\affiliation{Department of Physics, Princeton University, Princeton, NJ 08544, USA}
\author{J. M. Taylor}
\affiliation{Joint Quantum Institute/NIST, College Park, Maryland 20742, USA}
\author{A. A. Houck}
\affiliation{Department of Electrical Engineering, Princeton University, Princeton, New Jersey 08544, USA}
\author{J. R. Petta}
\affiliation{Department of Physics, Princeton University, Princeton, NJ 08544, USA}
\affiliation{Princeton Institute for the Science and Technology of Materials (PRISM), Princeton University, Princeton,
New Jersey 08544, USA}

\begin{abstract}
Circuit quantum electrodynamics allows spatially separated superconducting qubits to interact via a ``quantum bus", enabling two-qubit entanglement and the implementation of simple quantum algorithms. We combine the circuit quantum electrodynamics architecture with spin qubits by coupling an InAs nanowire double quantum dot to a superconducting cavity. We drive single spin rotations using electric dipole spin resonance and demonstrate that photons trapped in the cavity are sensitive to single spin dynamics. The hybrid quantum system allows measurements of the spin lifetime and the observation of coherent spin rotations. Our results demonstrate that a spin-cavity coupling strength of 1 MHz is feasible.
\end{abstract}


\maketitle
Electron spins trapped in quantum dots have been proposed as basic building blocks of a future quantum processor \cite{Loss1998,Hanson2007}. With spin qubits, two qubit operations are typically based on exchange coupling between nearest neighbor spins, leading to a fast 180 ps entangling gate \cite{Petta2005}. However, a scalable spin-based quantum computing architecture will almost certainly require long-range qubit interactions. Unfortunately, the weak magnetic moment of the electron makes it difficult to couple spin qubits that are separated by a large distance. Approaches to transferring spin information by physically shuttling electrons or using exchange-coupled spin chains are experimentally challenging \cite{Hermelin2011,McNeil2011,Friesen2007}. In comparison, circuit quantum electrodynamics (cQED) has enabled long distance coupling of multiple superconducting qubits via a microwave cavity, providing a scalable architecture for quantum computation \cite{Wallraff2004,Reed2012,Sillanpaa2007}. Several proposals suggest coupling spatially separated spin qubits via a microwave cavity, but direct coupling between a single spin and the magnetic field of the cavity results in a spin-cavity vacuum Rabi frequency $g_{\rm S}$/2$\pi$ $\sim$ 10 Hz; far too weak to be useful for quantum information processing \cite{Imamoglu2009,Childress2004}.

Here we harness spin-orbit coupling in a hybrid quantum dot/cQED architecture to couple the electric field of a high quality factor superconducting cavity to a single ``spin-orbit qubit" fabricated from an InAs nanowire double quantum dot (DQD) \cite{Wallraff2004,Schroer2011,NadjPerge2010,Fasth2007}. The architecture allows us to achieve a charge-cavity vacuum Rabi frequency $g_{\rm C}$/2$\pi$ $\sim$ 30 MHz, consistent with coupling rates obtained in GaAs quantum dots and carbon nanotubes \cite{Frey2012,Delbecq2011}. The strong spin-orbit interaction of InAs allows us to electrically drive spin rotations with a local gate electrode, while the 30 MHz cavity-charge interaction provides a measurement of the resulting spin dynamics. An alternative approach that has been recently explored consists of coupling ensembles of spins ($N$ $\sim$ 10$^{12}$) to superconducting resonators \cite{Schuster2010,Kubo2010,Amsuss2011,Zhu2011}.

Our hybrid spin-orbit qubit/superconducting device is shown in Fig.\ 1(a). We fabricate a $\lambda$/2 superconducting Nb resonator (the cavity) with a resonance frequency $f_{\rm 0}$ = $\omega_{\rm 0}$/2$\pi$ $\sim$ 6.2 GHz and quality factor, $Q$ $\sim$ 2000 \cite{SOM}. The amplitude and phase response of the cavity is detected using a homodyne measurement with a microwave probe frequency $f_R$ \cite{Wallraff2004}. We couple a single InAs nanowire spin-orbit qubit to the electric field generated by the cavity \cite{Trif2008}. The qubit consists of a DQD defined in an InAs nanowire \cite{Schroer2011,NadjPerge2010}. A series of Ti/Au depletion gates create a simple double well confinement potential containing ($N_{\rm L}$, $N_{\rm R}$) electrons, where $N_{\rm L}$ ($N_{\rm R}$) is the number of electrons in the left (right) dot. We tune the tunnel coupling, $t_{\rm C}$, of the DQD by adjusting the voltage $V_{\rm M}$ on the middle barrier gate, labeled M in Fig.\ 1(c). A trapped electron in the DQD has an electric dipole moment $d$ $\sim$ 1000 $e$$a_{\rm o}$, where $a_{\rm o}$ is the Bohr radius and $e$ is the electronic charge. To maximize the electric field at the position of the DQD, the drain contact of the nanowire is connected to the ground plane of the resonator and the source contact is connected to an anti-node of the resonator [see Fig.\ 1(b)]. Standard dc transport measurements are made possible by applying a source-drain bias, $V_{\rm SD}$, to the DQD via a spiral inductor that is connected to the voltage node of the resonator \cite{Chen2011}.

\begin{figure}
\begin{center}
		\includegraphics[width=0.7\columnwidth]{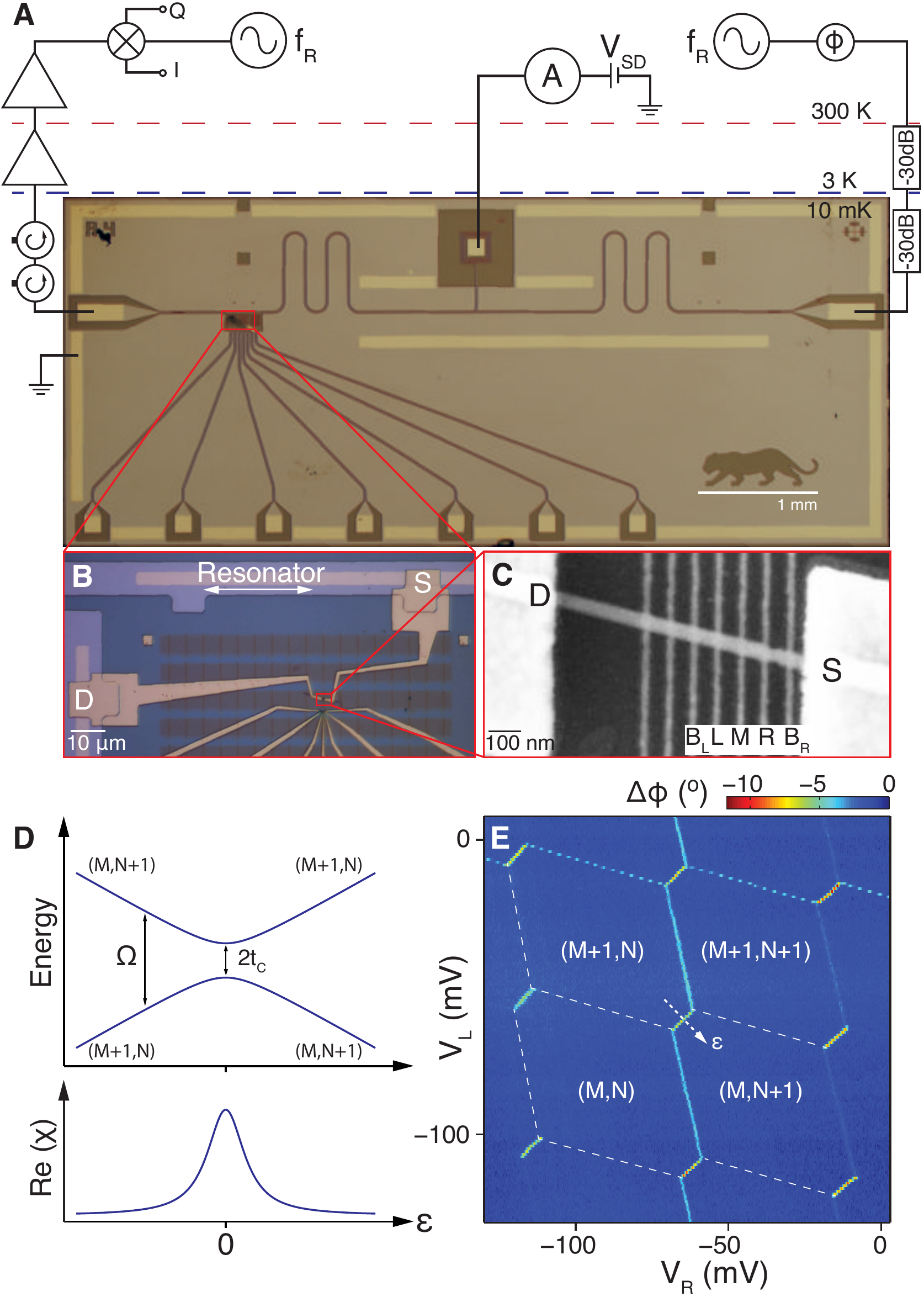}
\caption{\label{fig1} (Color online) (a) Circuit schematic and micrograph of a device similar to the one measured. Transmission through the $\lambda$/2 superconducting Nb resonator is measured using homodyne detection. Nanowire source-drain bias, $V_{\rm SD}$, is applied at the central voltage node of the cavity through a $\sim$ 4 nH inductor. (b) Expanded image of the region containing the DQD. One end of the nanowire is connected to the resonator ground plane and the other end is connected to the anti-node of the resonator. (c) Scanning electron micrograph image of the nanowire DQD. Seven gate electrodes are used to create a confinement potential along the length of the nanowire. (d) DQD energy levels (upper plot) and ac susceptibility, $\chi$, (lower plot) as a function of detuning, $\epsilon$. (e) The phase response of the resonator provides a direct measurement of the DQD charge stability diagram.}
\end{center}	
\vspace{-0.5cm}
\end{figure}

We focus on the cavity response near the ($M$, $N$+1) $\leftrightarrow$ ($M$+1, $N$) interdot charge transition. Neglecting spin for the moment, the DQD forms a two-level ``artificial molecule" with an energy splitting $\Omega = \sqrt{\epsilon^2+4t_{\rm C}^2}$, where $\epsilon$ is the detuning [upper diagram, Fig.\ 1(d)]. Interdot tunnel coupling hybridizes the charge states around $\epsilon$ = 0 resulting in a tunnel splitting of 2$t_{\rm C}$. The detuning dependent dipole moment of the DQD has an admittance which loads the cavity. We characterize the strength of the interaction by the ac susceptibility $\chi$ [lower panel, Fig. 1(d)] \cite{Blais2004}.

A qualitative understanding of the quantum dot/cavity coupling can be obtained considering the relevant energy scales in the system. The single dot charging energy, $E_{\rm C}$ $\sim$ 12 meV, is much larger than the relevant photon energies, $hf_{\rm R}$ $\sim$ 25  eV, and the cavity is largely unaffected by the DQD in Coulomb blockade. However, near interdot charge transitions (e.g. ($M$, $N$+1) $\leftrightarrow$ ($M$+1, $N$)), or transitions with the source and drain electrodes (e.g. ($M$, $N$)$\leftrightarrow$($M$, $N$+1)), the energy scales associated with the DQD are close to the cavity energy, and the cavity is damped, resulting in a negative phase shift in microwave transmission at the bare cavity frequency. The DQD charge stability diagram is measured in Fig.\ 1(e) by probing the phase response of the microwave cavity as a function of the gate voltages $V_{\rm R}$ and $V_{\rm L}$ \cite{Frey2012,Delbecq2011}.

Quantitative analysis of the cavity response requires a fully quantum mechanical model that accounts for photon exchange between the microwave field and the DQD \cite{Trif2008,Chen2011}. In cavity QED, one considers interactions between an atom with transition frequency $\omega_{\rm a} = \Omega/\hbar$ and the photon field of the cavity, characterized by the resonance frequency $\omega_{\rm 0}$. The atom and cavity energy levels hybridize when the atom-cavity detuning  $\Delta$ = $\omega_{\rm a}$ - $\omega_{\rm 0}$ $<$ $g_{\rm C}$, leading to the Jaynes-Cummings ladder of quantum states \cite{Jaynes1963}. When the atom and cavity are detuned in the dispersive limit ($\Delta$ $>$ $g_{\rm C}$), the cavity field exhibits a phase shift in microwave transmission at the bare cavity frequency that is given by $\phi$ = $\arctan[(2g_{\rm C}^2)/(\kappa \Delta)]$, where $\kappa$ is the cavity decay rate. The phase will therefore change sign as the atom-cavity detuning $\Delta$ is tuned from positive to negative values \cite{Wallraff2004}.

We extract $g_{\rm C}$ by measuring the amplitude and phase response of the cavity for several values of the interdot tunnel coupling [Figs.\ 2(e), (f)]. For example, with $V_{\rm M}$ = -2.26 V, the qubit transition frequency is always greater than the cavity frequency ($\Omega/\hbar$ $>$ $\omega_{\rm 0}$), leading to a negative phase shift for all values of $\epsilon$. In contrast, for $V_{\rm M}$ = -2.32 V, the minimum qubit transition frequency 2$t_{\rm C}$/$\hbar$ $<$ $\omega_{\rm 0}$ and $\Delta$ changes sign with $\epsilon$, resulting in a phase shift that takes on both positive and negative values. We fit the data to a master equation model using a best fit value of $g_{\rm C}$/2$\pi$  = 30 MHz and a $V_{\rm M}$ dependent tunnel coupling that ranges from 2$t_{\rm C}$/$h$= 1.8 to 7.0 GHz \cite{SOM}. We assume a qubit lifetime of 15 ns and account for inhomogeneous broadening due to charge noise by convolving the phase and magnitude response with a Gaussian of width  $\sigma_{\rm E}$ = 21 $\mu$eV \cite{Petersson2010b}. The vacuum Rabi frequency extracted here compares favorably to values obtained using Cooper pair box qubits, $g_{\rm C}$/2$\pi$ $\sim$ 6 MHz \cite{Wallraff2004},  transmon qubits, $g_{\rm C}$/2$\pi$ $\sim$ 100 MHz \cite{Schuster2007}, and many-electron GaAs quantum dots, $g_{\rm C}$/2$\pi$ $\sim$ 50 MHz \cite{Frey2012}.

\begin{figure}
\begin{center}
		\includegraphics[width=0.7\columnwidth]{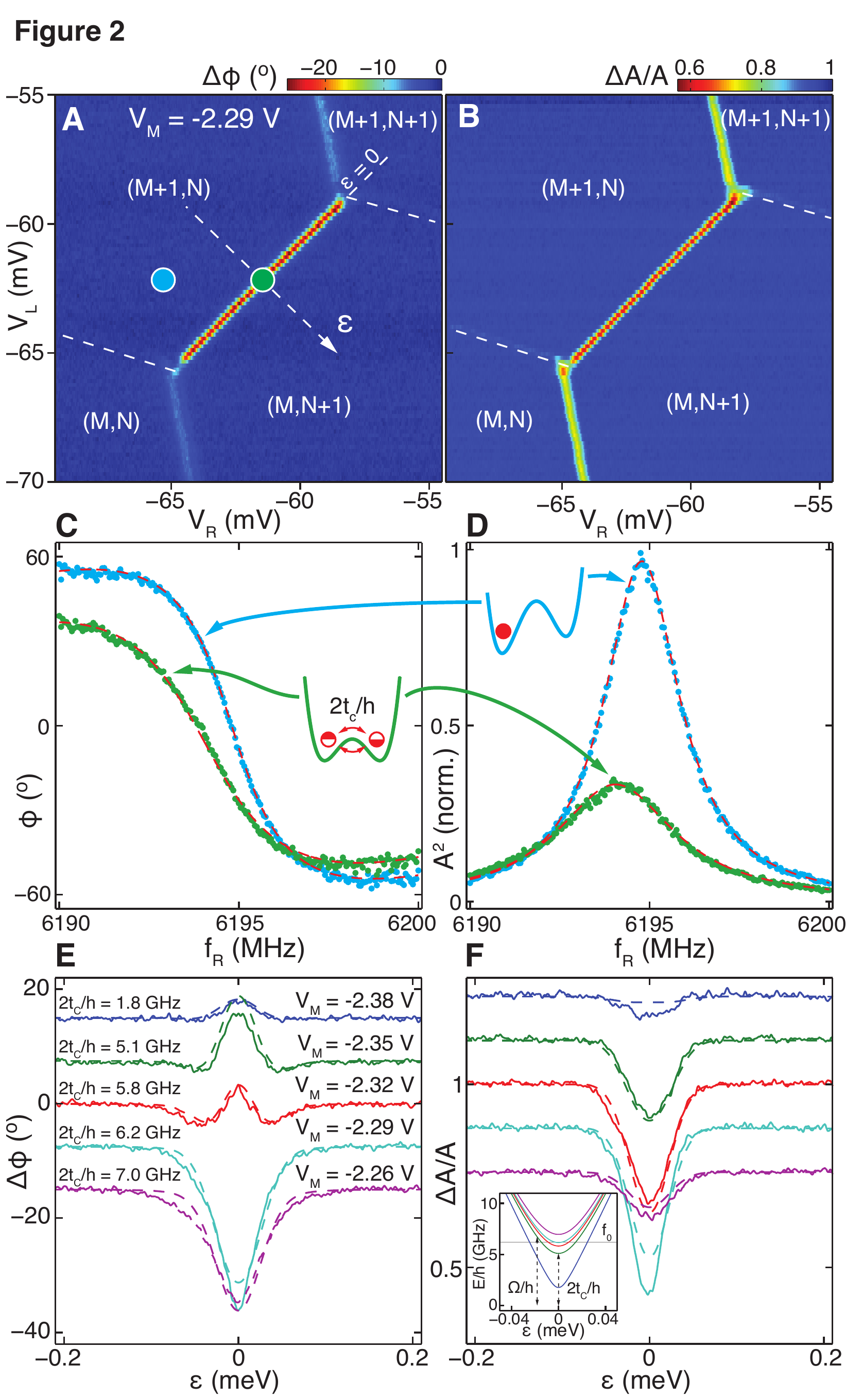}
\caption{\label{fig2} (Color online) (a) -- (b) Phase and amplitude response of the cavity near the ($M$+1, $N$) $\leftrightarrow$ ($M$, $N$+1) charge transition measured using a fixed drive frequency, $f_{\rm R}$ = 6194.8 MHz. (c) -- (d) Phase and normalized amplitude of the microwave field plotted as a function of $f_{\rm R}$ at the interdot charge transition (green curve) and in Coulomb blockade (blue curve). (e) -- (f) Phase and amplitude response measured as a function of DQD detuning, $\epsilon$, for a range of tunnel couplings, as set by $V_{\rm M}$. Dashed lines are fits to the data, allowing the extraction of the cavity coupling strength, $g_{\rm C}$ $\sim$ 30 MHz (see main text). Inset: Cavity frequency relative to the qubit transition frequency, $\Omega/h$.}
\end{center}	
\vspace{-0.6cm}
\end{figure}

We access the spin-degree of freedom by operating the device as a spin-orbit qubit (Fig.\ 3). For simplicity, we label the charge states (1,1) and (0,2) \cite{Petta2005}. The ground state with two-electrons in the right quantum dot is the singlet S(0,2). At negative detuning, the four relevant spin-orbital states are $|$$\Uparrow\Uparrow >$, $|$$\Downarrow\Downarrow>$, $|$$\Uparrow\Downarrow>$, and $|$$\Downarrow\Uparrow>$ \cite{NadjPerge2010}. The level diagram is similar to a GaAs singlet-triplet spin qubit, with a key difference being that the g-factors for the two spins can vary significantly \cite{Petta2005}. Interdot tunnel coupling hybridizes the states with singlet character near $\epsilon$ = 0, and an external field results in Zeeman splitting $E_{\rm Z}$ = $\tilde{g}\mu_{\rm B} B$ of the spin states, where $\tilde{g}$ is the electronic g-factor, $\mu_{\rm B}$ is the Bohr magneton, $B$ is the magnetic field.

Spin selection rules result in Pauli blockade at the two-electron transition, a key ingredient for spin preparation and measurement [see inset, Fig.\ 3(b)] \cite{Petta2005,NadjPerge2010,Ono2002}. For example, the $|$$\Uparrow\Uparrow>$ state cannot tunnel to S(0,2) due to Pauli exclusion. Modulation of the confinement potential with a gate voltage results in spin-orbit-driven EDSR transitions that lift the Pauli blockade \cite{NadjPerge2010,Golovach2006}. In Fig.\ 3(b) we plot the current, $I$, through the DQD with $V_{\rm SD}$ = 2.5 meV and the gates tuned in Pauli blockade [blue dot, inset Fig.\ 3(b)]. Hyperfine fields rapidly mix spin states when $E_{\rm Z}$ = $\tilde{g}\mu_{\rm B}B$ $<$ $B_{\rm N}$, where $B_{\rm N}$ $\sim$ 2 mT is the hyperfine field \cite{Schroer2011}. At finite fields, the leakage current is non-zero when the ac driving frequency on the gate, $f_{\rm G}$, satisfies the electron spin resonance condition $E_{\rm Z}$ = $h f_{\rm G}$. We observe two resonance conditions corresponding to single spin rotations in the left and right quantum dot, with g-factors of 8.2 and 10.6 \cite{NadjPerge2010}.

\begin{figure}
\begin{center}
	\includegraphics[width=0.75\columnwidth]{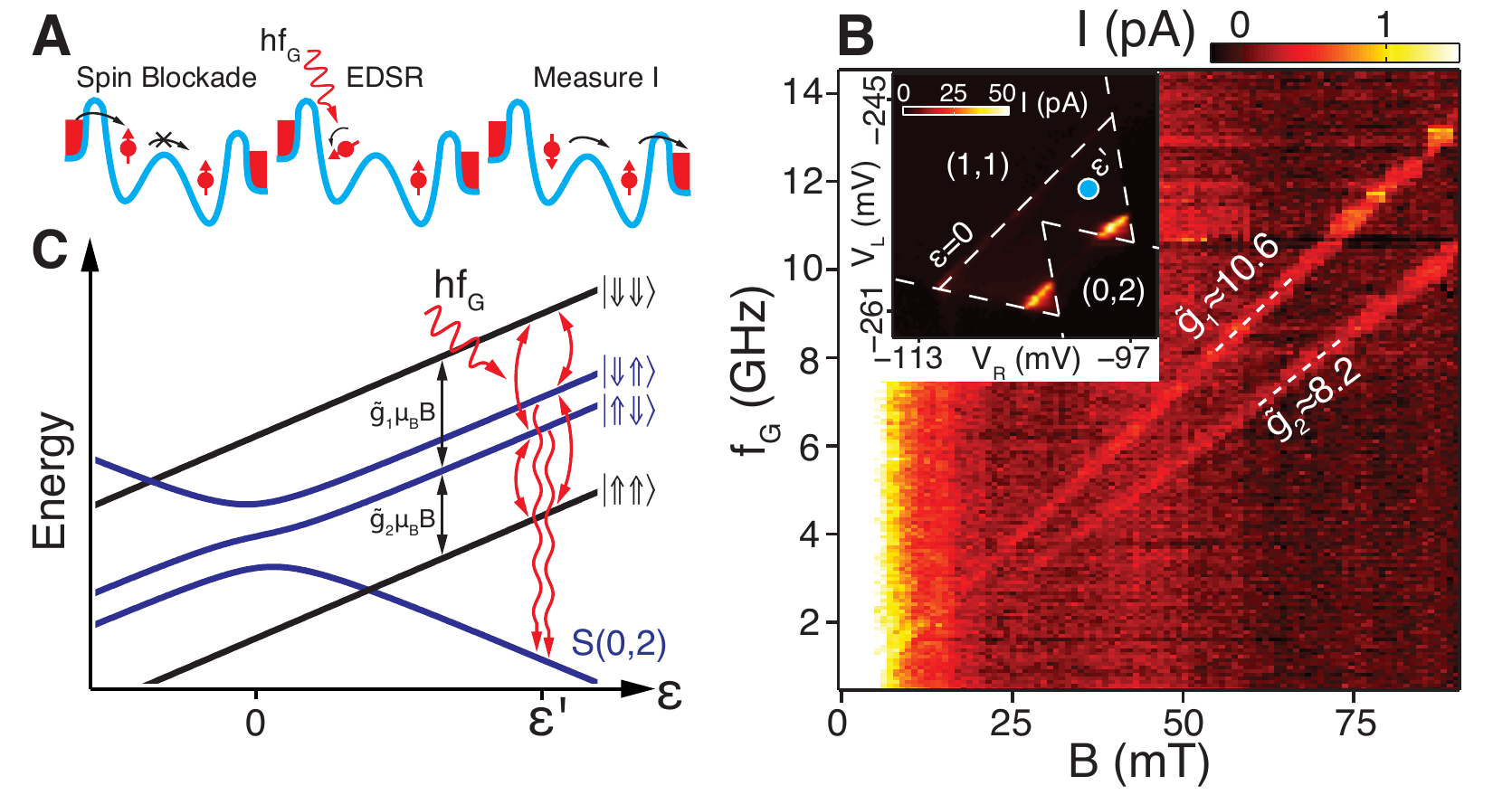}
\caption{\label{fig3} (Color online) (a) EDSR transitions lift Pauli blockade, resulting in current flow through the device. (b) Leakage current measured in Pauli blockade at point $\epsilon$$'$ (inset) as a function of magnetic field, $B$, and microwave driving frequency, $f_{\rm G}$. Pauli blockade is lifted by EDSR driving when $E_{\rm Z}$ = $\tilde{g} \mu_{\rm B} B$ = $h f_{\rm G}$. Inset: Finite-bias triangles measured with $V_{\rm SD}$ = 2.5 mV indicate a suppression of current due to Pauli blockade. (c) Energy levels of the spin-orbit qubit plotted as a function of $\epsilon$. The data in (b) are acquired with $\epsilon$ = $\epsilon$$'$.}
\end{center}	
\vspace{-0.5cm}
\end{figure}

Around $\epsilon$ = 0, the DQD has a spin state dependent dipole moment that allows spin state readout via the superconducting cavity \cite{Petersson2010a}. We combine quantum control of the spins using EDSR and cavity detection of single spin dynamics using the pulse sequence shown in Figs.\ 4 (a),(b). Starting with the spin qubit in state $|$$\Uparrow\Uparrow>$, we pulse to negative detuning and apply a microwave burst of length  $\tau_{\rm B}$ to drive EDSR transitions. For example, an EDSR $\pi$-pulse will drive a spin transition from $|$$\Uparrow\Uparrow>$ to $|$$\Uparrow\Downarrow>$. The resulting spin state is probed by pulsing back to $\epsilon$ = 0 for a time $T_{\rm M}$. The cavity is most sensitive to charge dynamics near $\epsilon$  = 0 due to the different ac susceptibility of the $|$$\Uparrow\Downarrow>$ and $|$$\Uparrow\Uparrow>$ spin states \cite{SOM}. In Fig.\ 4(c) we plot the cavity phase shift as a function of $f_{\rm G}$ and $B$. We again observe two features that follow the standard spin resonance condition, consistent with the dc transport data in Fig.\ 4(b). Varying the measurement time $T_{\rm M}$, we fit the measured phase response to an exponential decay and estimate a spin lifetime $T_{\rm 1}$ $\sim$ 1 $\mu$s [Fig.\ 4(d)].

We demonstrate time-resolved Rabi oscillations in the spin-orbit qubit and readout via the cavity by varying the EDSR microwave burst length $\tau_{\rm B}$. Figure 4(e) shows the measured phase as a function of $\tau_{\rm B}$ and gate drive power, $P_{\rm G}$. We observe Rabi oscillations with a minimum period of 17 ns, as shown in Fig.\ 4(f), consistent with an EDSR driving mechanism \cite{NadjPerge2010}. These results demonstrate that the microwave field of the cavity is sensitive to the spin state of a single electron.

\begin{figure}
\begin{center}
		\includegraphics[width=0.75\columnwidth]{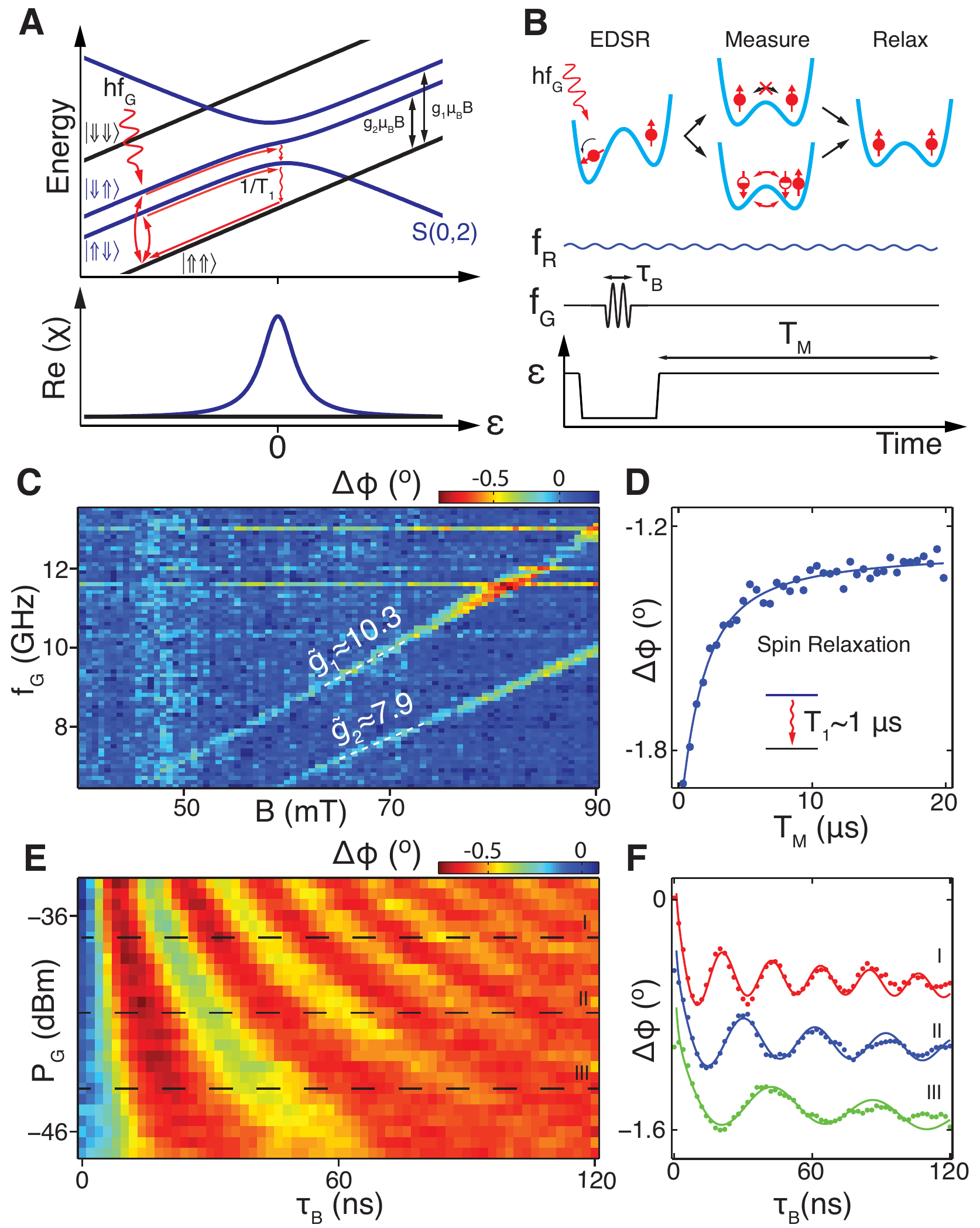}
\caption{\label{fig4} (Color online) (a) Upper: Pulse sequence used for resonator readout is superimposed on the level diagram. Lower: The ac susceptibility, $\chi$, is dependent on the spin state of the DQD and allows for sensitive spin readout. (b) Pulse sequence used to drive EDSR transitions. (c) Phase response of the cavity measured as a function of EDSR drive frequency, $f_{\rm G}$, and external field, $B$, with $\tau_{\rm B}$ = 100 ns and $T_{\rm M}$ = 850 ns. EDSR transitions are observed in the phase response, in agreement with the dc transport data. (d) Measured phase shift as a function of $T_{\rm M}$ with $\tau_{\rm B}$ = 100 ns, $B$ = 90 mT and $f_{\rm G}$ = 13.1 GHz. The exponential decay yields a spin relaxation time of $T_{\rm 1}$ = 1 $\mu$s. (e) Phase response of the cavity as a function of EDSR burst length, $\tau_{\rm B}$, and driving power for fixed $B$ = 86 mT, $f_{\rm G}$ = 9.5 GHz, and $T_{\rm M}$ = 1.75 $\mu$s. (f) Rabi oscillations at different powers, indicated by the dashed lines (e). The solid curves are fits to a power law decay \cite{SOM}.}
\end{center}	
\vspace{-0.6cm}
\end{figure}

In cQED, a large number of qubits can be connected via the electric field of the superconducting cavity. Trif \textit{et al.} have proposed using the cQED approach to couple two spin-orbit qubits via a cavity mediated interaction \cite{Trif2008}. Based on our results, we can estimate the effective spin-cavity coupling strength using theory developed for single quantum dots, $g_{\rm S} \approx g_{\rm C} \dfrac{E_{\rm Z}}{\Delta E_{\rm 0}}\dfrac{R}{\lambda_{\rm SO}}$ \cite{Trif2008}. Taking $g_{\rm C}$/2$\pi$  = 100 MHz (which can be obtained by increasing the cavity frequency), $E_{\rm Z}$ = 70 $\mu$eV, an orbital level spacing  $E_{\rm O}$ = 1.7 meV, dot radius $R$ = 25 nm, and spin-orbit length $\lambda_{\rm SO}$ = 100--200 nm, we find a spin coupling rate $g_{\rm S}$/2$\pi$ $\sim$ 1 MHz, which is five orders of magnitude larger than the coupling rate that would be obtained by coupling a single spin to the magnetic field of a microwave cavity.

In order to implement coherent state transfer between the qubit and the cavity the device must be in the strong coupling regime, where the spin-cavity coupling rate $g_{\rm S}$ is  larger than the cavity decay rate $\kappa$ and the qubit decoherence rate $\gamma$. Optimization of the resonator design will reduce the cavity decay rate to well below 1 MHz \cite{SOM}. There are several options for decreasing the qubit decoherence rate. First, dynamical decoupling can be used to reduce the qubit decay rate to $\sim$ 1 MHz in the InAs system \cite{NadjPerge2010}. InAs could also be replaced by Ge/Si core/shell nanowires where hole spin-orbit coupling is predicted to be large \cite{Kloeffel2011}. Resonators can also be coupled to nuclear-spin-free Si/SiGe quantum dots by using micromagnets to create artificial spin-orbit fields \cite{PioroLadriere2008}. Based on our results we anticipate that the strong coupling regime for single spins can be reached, eventually allowing spin qubits to be interconnected in a quantum bus architecture.

Acknowledgements: Research at Princeton was supported by the Sloan and Packard Foundations, Army Research Office grant W911NF-08-1-0189, DARPA QuEST award HR0011-09-1-0007 and the National Science Foundation through the Princeton Center for Complex Materials, DMR-0819860, and CAREER award, DMR-0846341. JMT acknowledges support from ARO MURI award W911NF0910406.

\end{document}